\documentclass[twocolumn, aps, prb, amsmath, 10pt, floatfix]{revtex4-2}

\usepackage[T1]{fontenc}
\usepackage[utf8]{inputenc}

\usepackage{amsmath}
\usepackage{dsfont}
\usepackage{mathrsfs}
\usepackage{braket}
\usepackage{paralist}
\usepackage{afterpage}
\usepackage{graphicx}
\graphicspath{ {./figures/} }

\usepackage{soul}

\begin{document}

\title{Optimal Control of the Operating Regime of a Single Electron Double Quantum Dot}

\author{Vincent Reiher}
\author{Yves B\'{e}rub\'{e}-Lauzi\`{e}re}
\email{Yves.Berube-Lauziere@USherbrooke.ca}
\affiliation{D\'{e}partement de g\'{e}nie \'{e}lectrique et de g\'{e}nie informatique}
\affiliation{Institut quantique, Universit\'{e} de Sherbrooke, Sherbrooke, Qu\'{e}bec J1K 2R1, Canada}

\begin{abstract}
The double quantum dot device benefits from the advantages of both the spin and charge qubits, while offering ways to mitigate their drawbacks. Careful gate voltage modulation can grant greater spinlike or chargelike dynamics to the device, yielding long coherence times with the former and high electrical susceptibility with the latter for electrically driven spin rotations or coherent interactions with microwave photons. We show that optimal control pulses generated using the \textit{GRadient Ascent Pulse Engineering} (GRAPE) algorithm can yield higher-fidelity operating regime transfers than can be achieved using linear methods.
\end{abstract}

\maketitle

\section{Introduction}
\label{sec:intro}
Ever since Loss and DiVincenzo's proposal to use the electron spin as the fundamental building block for quantum computing~\cite{Loss1998}, important efforts have been devoted to the development of spin qubit architectures~\cite{Burkard2000, Hanson2007, Mills2019, Sigillito2019-1}. The long coherence times of electron spins, which have reached the order of seconds in silicon~\cite{Hanson2007, Tyryshkin2012, Zwanenburg2013, Veldhorst2014}, as well as the existing fabrication infrastructure of the silicon industry, make them great candidates for physical qubit implementations. However, although high-fidelity initialization, manipulation and readout of small numbers of spins isolated within quantum dots has been demonstrated~\cite{Mills2019, Sigillito2019-2, Zajac2018, Yang2019, Meunier2011, Watson2018}, the fabrication of two-dimensional arrays of interconnected spins, required for quantum information processing and error correction, remains an outstanding challenge~\cite{Chanrion2020, Mortemousque2018, Hendrickx2020}.

One avenue to solving this challenge lies in using microwave photons in superconducting resonators to mediate long-range spin-spin interactions, as has been demonstrated for superconducting qubits~\cite{Blais2004, Wallraff2004, DiCarlo2009}. Coherent interactions between single spins and microwave photons have already been shown using the large electric dipole of the electron charge state in a double quantum dot (DQD) through spin-charge hybridization~\citep{Benito2017, Mi2018, Viennot2015}. Conversely, the increased electrical susceptibility of such a device can be used to drive spin state rotations via electric dipole spin resonance (EDSR) by quickly displacing the electron wave function in a local transverse magnetic field gradient~\cite{PioroLadriere2007, Yoneda2015, Benito2019, Croot2020}. Additionally, gate voltages can be modulated to reshape the double-well potential and reach the single dot regime, decoupling the spin and charge degrees of freedom and recovering long coherence times~\citep{Benito2017, Benito2019}. 
 
The single electron DQD therefore presents itself as a promising architecture for quantum computation, with two outstanding operating regimes:
\begin{compactitem}
\item A memory-mode regime wherein the electron is strongly localized in one of the two wells of the DQD due to a large interdot energy detuning. In this regime, the qubit dynamics approach that of a pure spin and are largely decoupled from environmental charge noise, recovering the long natural coherence times of electron spins in silicon;
\item A flopping-mode regime corresponding to a set of configurations wherein the electron charge state is delocalized across the two dots of the DQD. When the electron orbital and Zeeman energies approach resonance, this regime allows fast manipulation of the electron spin state via EDSR to realize one-qubit gates, or reaching the strong coupling regime between the electron spin and a microwave photon for long-range interactions between distant spins or for dispersive measurement of the qubit state.
\end{compactitem}

In the context of a quantum computation, such a device will be transferred many times between these operating regimes. It is therefore crucial that this transfer be made quickly and that it preserves maximal state fidelity in the logical basis. This is the problem dealt with in this paper. The GRAPE algorithm~\citep{Khaneja2005} allows fast generation of optimal control signals which have been used experimentally to manipulate physical qubits~\citep{Nimbalkar2012, Groszkowski2011, Yang2020, Zong2021}. GRAPE is also much easier to work with than traditional techniques relying on optimal control theory directly~\citep{Accanto2017, Castelano2018, Rasanen2013, Coden2019}. It will be demonstrated that control signals obtained using the GRAPE algorithm allow faster and higher-fidelity operating regime transfers than can be attained using standard linear approaches. This is achieved independently of the qubit's state and therefore does not require a priori knowledge of the qubit state, which is of prime importance in practice.

The remainder of this paper is organized as follows. Section \ref{sec:model} describes the model of the DQD device considered herein. The control objective used to optimize control pulses to alter the device's operating regime while preserving its logical state is formulated in Section \ref{sec:ctrl_obj}. Section \ref{sec:results} provides the results obtained from the application of optimal control pulses to transfer between the different operating regimes of the system for various qubit states. Finally, Section \ref{sec:experiment} discusses considerations for the experimental implementation of the resulting optimal control pulses. 

\section{Model}
\label{sec:model}

\begin{figure}[tbp]
\includegraphics[width=3.375in]{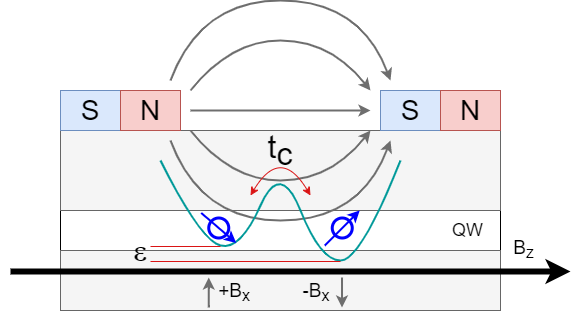}
\caption{Schematic of the device under control. A single electron is trapped within a double well potential. The device is influenced by an external magnetic field $B_z$, a local magnetic field gradient $B_x$ and gate voltages controlling the detuning energy $\epsilon$ and tunnel coupling $t_c$. Figure adapted from~\citep{Benito2017}.}
\label{fig:model}
\end{figure}

The physical system consists of a gate-defined DQD in silicon (Fig. \ref{fig:model}), tuned to the single electron regime, as described in Benito \textit{et. al}~\cite{Benito2017}. The device is subjected to a strong homogeneous longitudinal magnetic field $B_z$, and a weak transverse magnetic field gradient $B_x$. Gate voltages directly control the tunnel coupling $t_c$ between the left and right dot as well as the energy detuning $\epsilon$ between the two dots. The DQD system Hamiltonian can be written as
\begin{equation}
\label{eq:Hamiltonian}
H = \frac{\hbar}{2} ( \epsilon \tau_z + 2t_c \tau_x + g \mu_B B_z \sigma_z + g \mu_B B_x \sigma_x \tau_z ),
\end{equation}
where $\tau_\alpha$ are the Pauli matrices for the charge degree of freedom, in the left-right basis, and $\sigma_\alpha$ are the Pauli matrices for the spin degree of freedom. The valley degree of freedom is neglected here, as recent measurements on similar devices have shown sufficiently large valley splittings~\cite{Chen2021}. The eigenstates of this four level Hamiltonian are written as~\citep{Benito2017}
\begin{align}
\label{eq:H-state0}
\ket{0} &\approx \ket{-,\downarrow}, 	\\
\label{eq:H-state1}
\ket{1} &= \cos \frac{\phi}{2} \ket{-,\uparrow} + \sin \frac{\phi}{2} \ket{+,\downarrow},	\\
\ket{2} &= - \sin \frac{\phi}{2} \ket{-,\uparrow} + \cos \frac{\phi}{2} \ket{+,\downarrow},	\\
\ket{3} &\approx \ket{+,\uparrow},
\end{align}
with $\phi$, the spin-orbit mixing angle. The symmetric and anti-symmetric charge states are defined as
\begin{align}
\ket{+} &= \cos \frac{\theta}{2} \ket{L} + \sin \frac{\theta}{2} \ket{R},    \\
\ket{-} &= -\sin \frac{\theta}{2} \ket{L} + \cos \frac{\theta}{2} \ket{R},
\end{align}
with $\ket{L}$, $\ket{R}$ the electron charge states corresponding to occupation in the left and right dot respectively, and $\theta = \frac{\pi}{2} - \arctan \frac{\epsilon}{2t_c}$, the orbital angle. For the purposes of the present work, the qubit is defined on the $\ket{0} \leftrightarrow \ket{1}$ transition. The spin-orbit mixing angle describes the spinlike or chargelike character of the qubit: it is readily seen from Eqs. (\ref{eq:H-state0}) and (\ref{eq:H-state1}) that for $\phi = 0$, this transition forms a pure spin qubit, and that for $\phi = \pi$, this transition forms a pure charge qubit. The spin-orbit mixing angle is given by 
\begin{equation}
\phi = \arctan \frac{g\mu_B B_x \cos \theta}{\Omega - g\mu_B B_z},
\end{equation} 
where $\Omega = \sqrt{\epsilon^2 + 4t_c^2}$ is the orbital energy. Due to the transverse magnetic field gradient, the electron dipole operator acquires off-diagonal elements which couple the $\ket{-,\uparrow}$ and $\ket{+,\downarrow}$ states, leading to anticrossings in the energy spectrum as the detuning energy $\epsilon$ varies~\citep{Benito2017}. 

In addition to coherent evolution according to its Hamiltonian, the DQD is in general subject to dephasing and relaxation in both the spin and charge subspaces. Direct spin relaxation is typically a very slow process~\citep{Hayes2009}, and is neglected in this work. Spin dephasing via hyperfine interactions and charge dephasing due to quasistatic charge noise in the device are stable processes with rates which do not vary appreciably over the duration of a control sequence; they can therefore be approximated to constant rates. For the purpose of numerical simulation of decoherence processes in this work, the charge and spin dephasing rates are set to $\gamma_{\phi,c}/{2\pi} = 36$~MHz and $\gamma_{\phi,s}/{2\pi} = 1.2$~MHz, respectively, according to literature~\citep{Benito2017, Mi2018}. Finally, charge relaxation via emission of longitudinal acoustic (LA) phonons in the silicon lattice is the dominating process and is dependent on the control amplitudes $\epsilon$ and $t_c$ via the orbital energy $\Omega \equiv \Omega (\epsilon, t_c)$. This rate has been set to a constant $\gamma_{1,c} / 2\pi = 45$~MHz in this work, given the dynamical range considered for parameters $\epsilon$ and $t_c$, in an effort to reduce the computational complexity of the problem and avoid transcendental equations (see Appendix \ref{app:charge_relax}). While this noise model for the device is only an approximation of the real processes which would be experimentally measured, an exact description of the decoherence channels for the DQD is outside the scope of this work and is not expected to appreciably impact the results presented below, as the control operation timescales considered here are very short relative to typical decoherence rates for this type of device.

\section{Control Objective}
\label{sec:ctrl_obj}
The aim is to transfer the qubit between an initial and a final operating regime while preserving the qubit's logical state. Each operating regime is defined by its spin-orbit mixing angle $\phi$ or by its orbital energy $\Omega$ and orbital angle $\theta$. These parameters in turn define values for the inter-dot energy detuning $\epsilon$ and tunnel coupling $t_c$, which are the directly electrically controlled parameters.

The control objective is therefore to preserve the initial logical state. Let the initial qubit logical state be written in general form as
\begin{equation}
\ket{\Psi_i} = \alpha_i \ket{0_i} + \beta_i \ket{1_i},
\end{equation}
where $\ket{0_i}$, $\ket{1_i}$ are eigenstates of the DQD Hamiltonian in the initial operating regime. Using the initial and final Hamiltonian eigenbases as logical bases
\begin{equation}
\mathscr{L}_{i,f} = \lbrace \ket{0_{i,f}}, \ket{1_{i,f}}, \ket{2_{i,f}}, \ket{3_{i,f}} \rbrace,
\end{equation}
the column vectors containing the coefficients of the initial and final logical qubit states which represent these states with respect to the initial and final eigenbases are written as
\begin{align}
[\Psi_i]_{\mathscr{L}_i} &= \begin{bmatrix}
\alpha_i & \beta_i & 0 & 0
\end{bmatrix}^T,    \\
[\Psi_f]_{\mathscr{L}_f} &= \begin{bmatrix}
\alpha_f & \beta_f & \gamma_f & \delta_f
\end{bmatrix}^T.
\end{align}
However, in order to preserve the qubit's logical state through the operating regime transfer, it is necessary that the total effect of the control in the logical basis be equal to the identity; that is,
\begin{equation}
[\Psi_f]_{\mathscr{L}_f} = \mathds{1}_{\mathscr{L}_f\mathscr{L}_i} [\Psi_i]_{\mathscr{L}_i} = \begin{bmatrix}
\alpha_i & \beta_i & 0 & 0
\end{bmatrix}^T,
\label{eq:ctrl_identity_cond}
\end{equation}
where $\mathds{1}_{\mathscr{L}_f \mathscr{L}_i}$ is an identity matrix with ones on its diagonal. The coefficients of the logical qubit state in the initial eigenbasis are therefore preserved, and population of the higher excited states is kept to $\gamma_f = \delta_f = 0$, protecting the quantum information encoded within the qubit state.

Considering instead the effect of the control in the product basis of the DQD charge and spin states, henceforth referred to as the physical basis,
\begin{equation}
\mathscr{P} = \lbrace \ket{L}, \ket{R} \rbrace \otimes \lbrace \ket{\downarrow}, \ket{\uparrow} \rbrace,
\end{equation}
the ideal evolution of the initial physical state to the final physical state in this basis, in the absence of decoherence, is written as
\begin{equation}
[\Psi_f]_{\mathscr{P}} = \mathbf{U}_{\mathscr{PP}} [\Psi_i]_{\mathscr{P}},
\end{equation}
with $\mathbf{U}_{\mathscr{PP}}$, the evolution operator expressed in the physical basis and obtained by the usual exponentiation of the integral of the Hamiltonian operator over the evolution time. The exact expression is not important here, as will be seen next; this is, however, the operator which must be synthesized via optimal control. Using the basis change matrix $\mathcal{P}_{\mathscr{P} \leftarrow \mathscr{L}_{i,f}}$ from the initial or final logical basis to the physical basis, the total effect of the operating regime transfer in the logical basis is given by
\begin{equation}
[\Psi_f]_{\mathscr{L}_f} = \mathcal{P}_{\mathscr{L}_f \leftarrow \mathscr{P}} \mathbf{U}_\mathscr{PP} \mathcal{P}_{\mathscr{P} \leftarrow \mathscr{L}_i} [\Psi_i]_{\mathscr{L}_i}.
\label{eq:logic_basis_evo_op}
\end{equation}
Recall that $\mathcal{P}_{\mathscr{P} \leftarrow \mathscr{L}_i}$ contains in its columns the coefficients of the basis vectors of $\mathscr{L}_i$ decomposed along the basis vectors of $\mathscr{P}$ (the notation in~\citep{Lay2014} is used here). Identification of Eq. (\ref{eq:logic_basis_evo_op}) with Eq. (\ref{eq:ctrl_identity_cond}) as a condition on the effect of the control yields
\begin{equation}
\mathds{1}_{\mathscr{L}_f\mathscr{L}_i} = \mathcal{P}_{\mathscr{L}_f \leftarrow \mathscr{P}} \mathbf{U}_\mathscr{PP} \mathcal{P}_{\mathscr{P} \leftarrow \mathscr{L}_i},
\end{equation} 
which in turn leads to an analytical expression for the target control operator
\begin{equation}
\mathbf{U}_\mathscr{PP} = \mathcal{P}_{\mathscr{P} \leftarrow \mathscr{L}_f} \mathcal{P}_{\mathscr{L}_i \leftarrow \mathscr{P}}.
\end{equation}
It is this operator towards which the optimization algorithm must converge.

\begin{figure}[!tbp]
\includegraphics[width=3.375in]{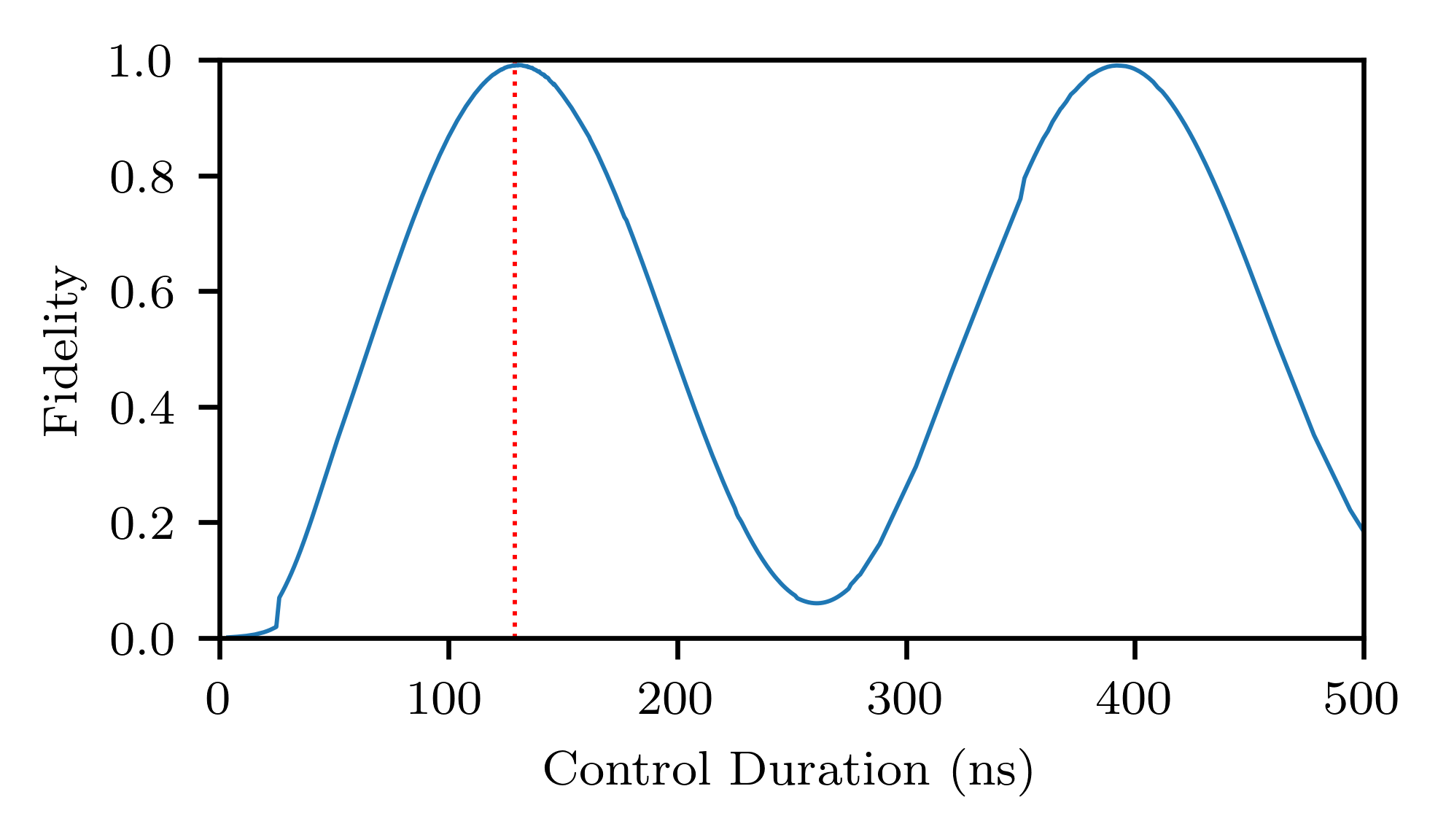}
\caption{Optimized control operator fidelity as a function of control pulse duration $T$. This particular simulation aims to design the pulse required to transfer the device from the spin qubit configuration to the flopping-mode configuration. Peak control operator fidelity of $99\%$ is first reached for a $129$~ns control pulse duration (vertical dotted line).}
\label{fig:op_fid}
\end{figure}

The following time-dependent Hamiltonian is used as a starting point for optimization: 
\begin{equation}
H(t) = \frac{\hbar}{2} ( \epsilon(t) \tau_z + 2t_c(t) \tau_x + g \mu_B B_z \sigma_z + g \mu_B B_x \sigma_x \tau_z),
\end{equation}
along with the controls
\begin{align}
&\epsilon(t) = \epsilon_i + \frac{t}{T}(\epsilon_f - \epsilon_i) + u_\epsilon (t),    \\
&t_c(t) = t_{c_i} + \frac{t}{T} (t_{c_f} - t_{c_i}) + u_{t_c} (t),
\end{align}
where $\epsilon_{\lbrace i, f \rbrace}$ and $t_{c_{\lbrace i, f \rbrace}}$ are the initial and final detuning and tunnel coupling energies and $T$ is the control pulse duration. It is seen that each of these controls consists of a linearly ramped part (which is customarily used) plus an additional control term $u_\epsilon (t)$ or $u_{t_c}(t)$. These additional terms are here iteratively optimized using the GRAPE~\cite{Khaneja2005} algorithm such that the operator representing the overall effect of the controls converges towards operator $\mathbf{U}$. Due to uncontrollable Hamiltonian evolution of the spin degree of freedom, the operator fidelity obtained with optimal control pulses exhibits an oscillating behavior with regard to the duration of the control sequence and it is expected that high fidelity operating regime transfers can only be performed for some precise pulse durations; see Fig. \ref{fig:op_fid}.

\begin{figure}[!tbp]
\includegraphics[width=3.375in]{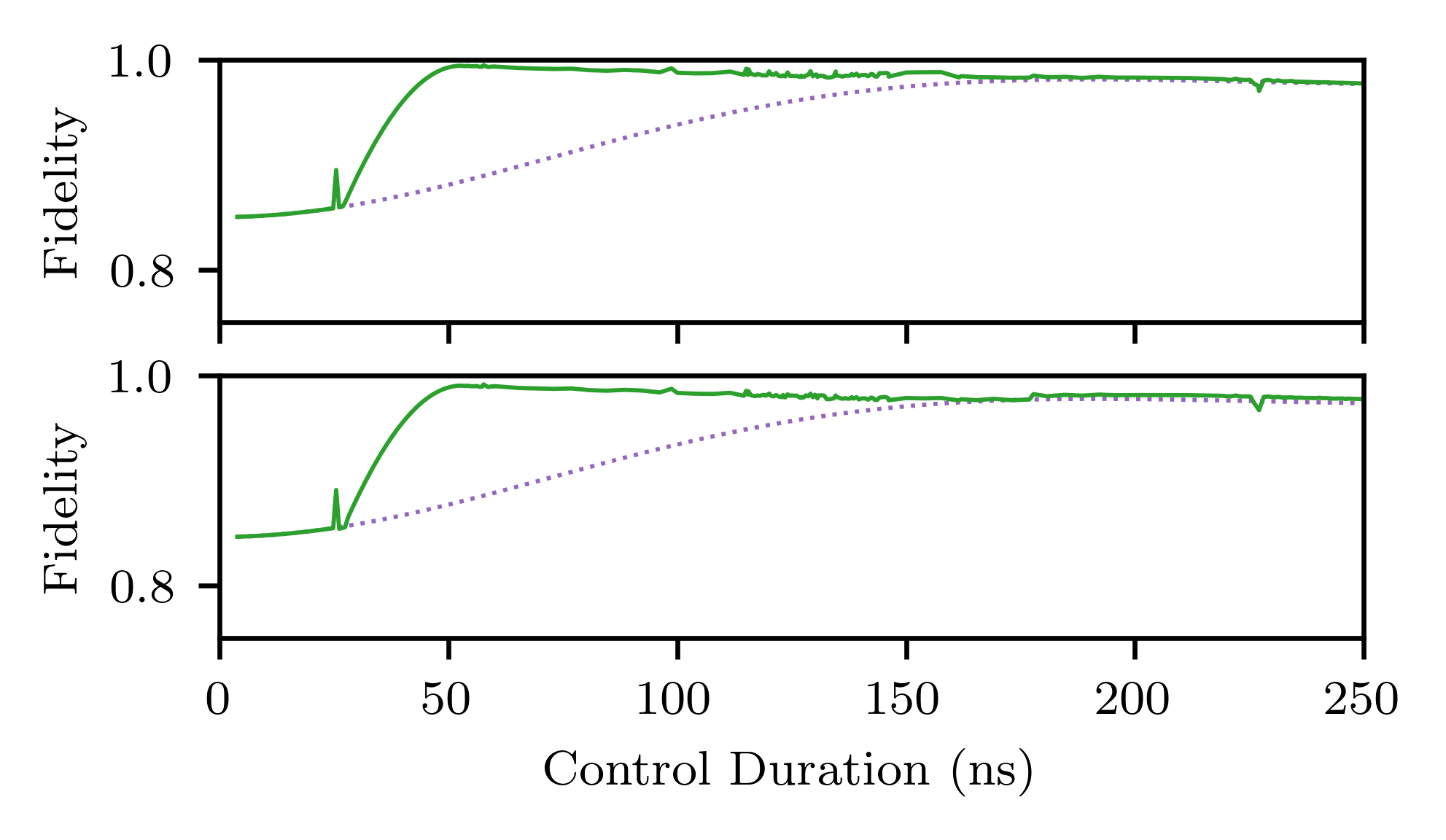}
\caption{Operating regime transfer for qubit eigenstates $\ket{0}$ (top) and $\ket{1}$ (bottom), from the memory-mode configuration to the flopping-mode configuration, as a function of control pulse duration $T$. In both cases, peak controlled fidelity greater than $99\%$ is reached for a $52.8$~ns pulse duration (solid line), whereas linear ramping of the electrical parameters (dotted line) reaches a peak fidelity of $97.7\%$ for $193$~ns pulses.}
\label{fig:eigen_tx_fid}
\end{figure}

\section{Numerical Results}
\label{sec:results}

The optimized control pulses are now applied to noisy evolution of the device model to simulate the operating regime transfer from the memory-mode spin qubit configuration to the flopping-mode qubit configuration with strong chargelike dynamics. The memory-mode qubit operating regime corresponds to a strongly detuned double well potential with $\epsilon = 40$~$\mu$eV, $t_c = 10$~$\mu$eV, whereas the flopping-mode configuration is defined as $\epsilon = 0$~$\mu$eV, $t_c = 16$ $\mu$eV. This symmetric configuration increases the spin-charge hybridization, yielding strong coupling of the electron spin to the electric field while keeping the added charge noise sensitivity to a minimum, as discussed in~\citep{Benito2017}. The static magnetic fields used are $g\mu_B B_x = 1.62$~$\mu$eV and $g\mu_B B_z = 24$~$\mu$eV.

\subsection{Qubit Eigenstate Transfer}
\label{sec:eigen_tx}

When the qubit state is initially prepared in an eigenstate ($\ket{\Psi_i} = \ket{0_i}$ or $\ket{\Psi_i} = \ket{1_i}$), it is found that the optimal control pulses generated previously reach high-fidelity operating regime transfers much faster than regular linear ramping of the electrical parameters $\epsilon$ and $t_c$ (Fig. \ref{fig:eigen_tx_fid}). Using GRAPE-designed control pulses represents a $72\%$ reduction in the time required to perform this operating regime transfer while also allowing higher-fidelity preservation of the qubit state.
\begin{figure}[!tbp]
\includegraphics[width=3.375in]{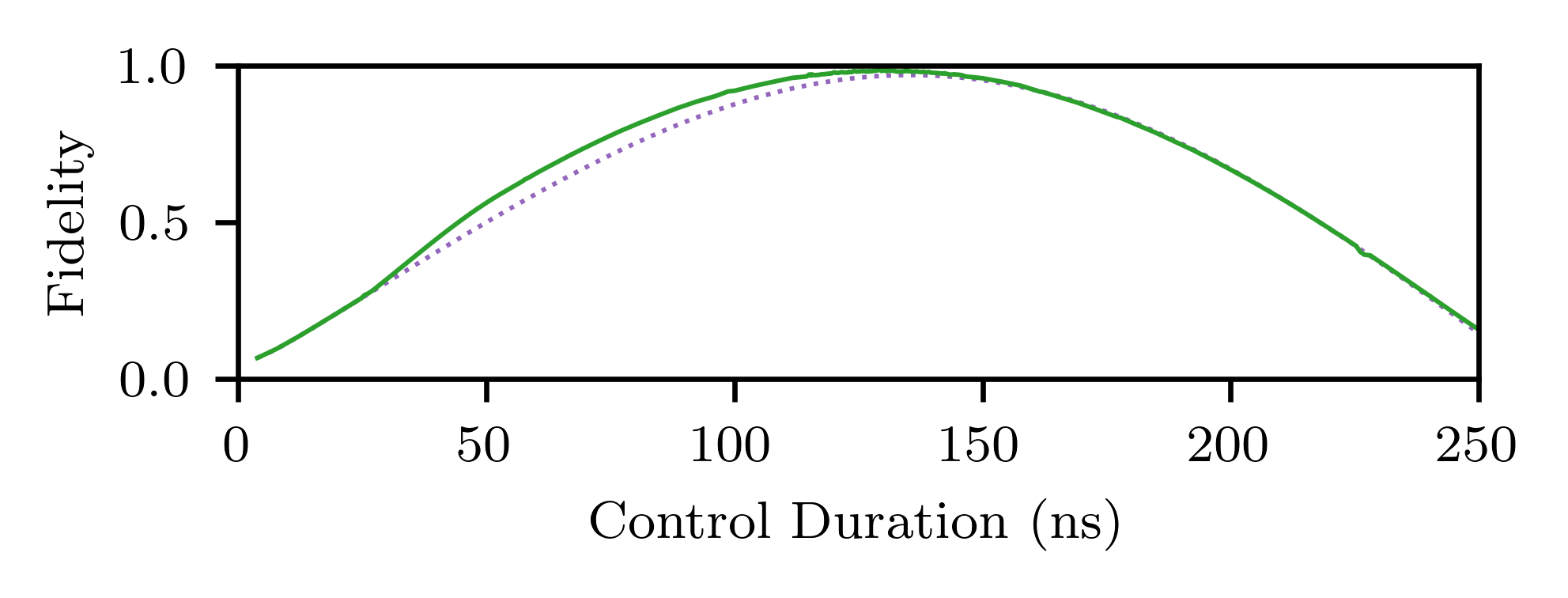}
\caption{Operating regime transfer operation for non-eigenstate $\ket{0} + \ket{1}$, from the memory-mode configuration to the flopping-mode configuration, as a function of control pulse duration $T$.  A $134.6$~ns optimal control pulse (solid line) reaches a fidelity of $98.6\%$,  whereas the linear ramp (dotted line) reaches a maximal fidelity of $96.9\%$.}
\label{fig:arbitrary_tx_fid}
\end{figure}
\begin{figure}[!tbp]
\includegraphics[width=3.375in]{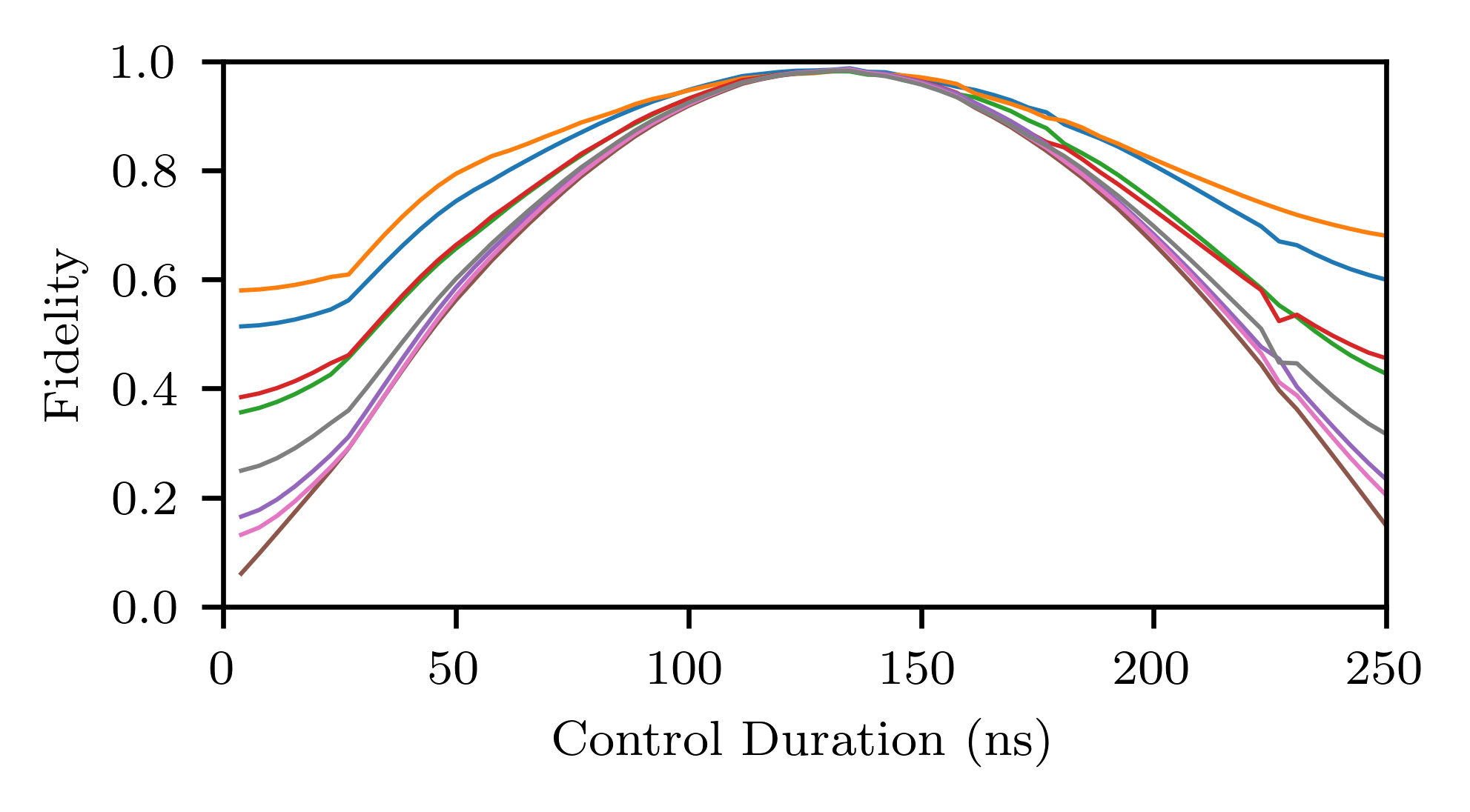}
\caption{Operating regime transfer operation for eight randomly-generated logical qubit states. GRAPE-generated optimal control pulses reach a peak fidelity greater than $99\%$ for $134.6$ ns pulses regardless of the logical qubit state.}
\label{fig:random_tx_fid}
\end{figure}
\subsection{Arbitrary Qubit State Transfer}
\label{sec:arbitrary_tx}
\begin{figure}[!tbp]
\includegraphics[width=3.375in]{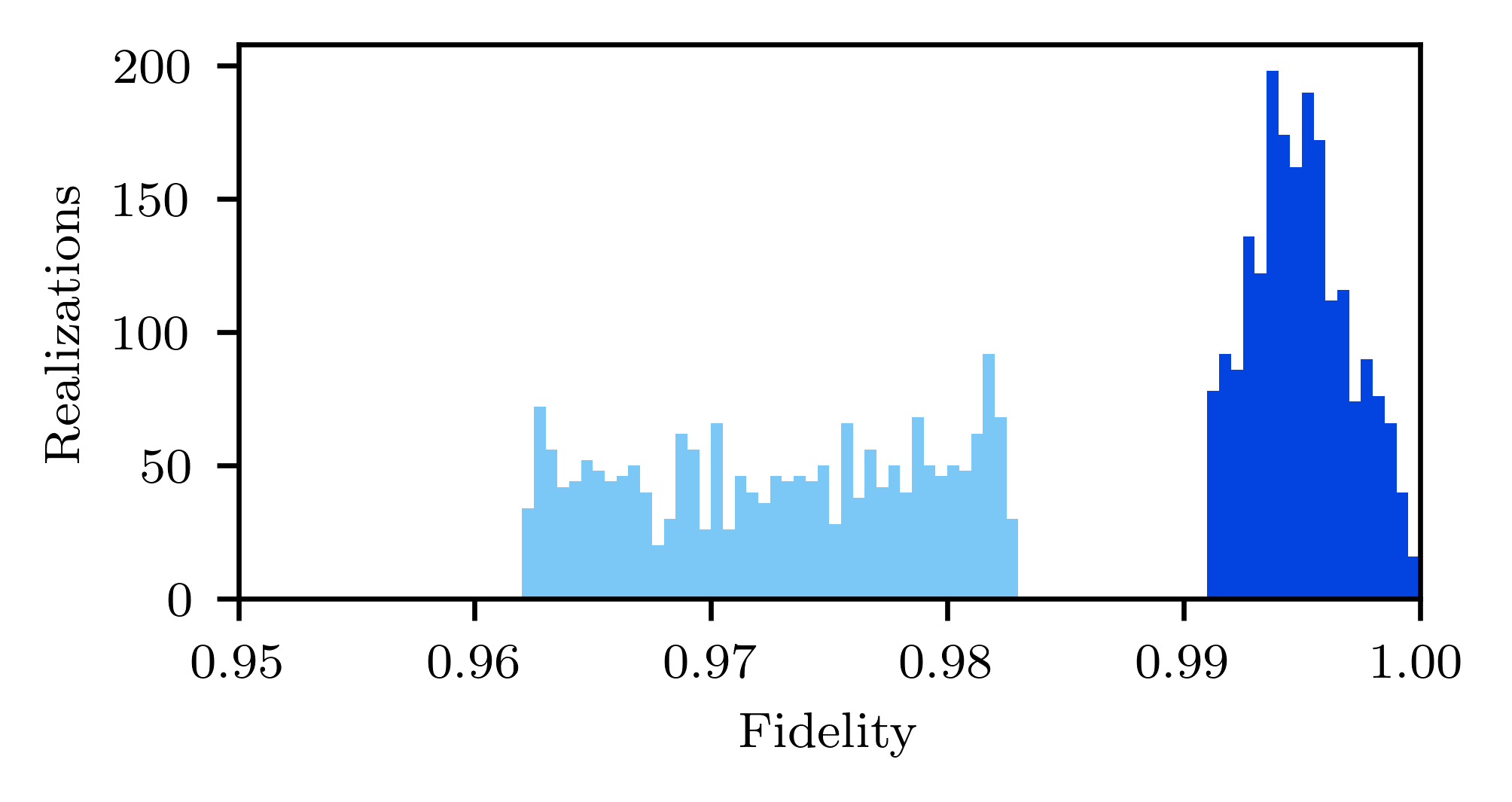}
\caption{Evolution statistics for $1000$ randomly-generated logical states. The GRAPE-optimized $134.6$~ns control pulse (dark blue) leads to a $2.2\%$ increase in average state fidelity, corresponding to a $81.3\%$ reduction in average error, and a $67\%$ reduction in standard deviation over linear ramping of the electrical parameters over the same duration (light blue).}
\label{fig:random_tx_stats}
\end{figure}
\begin{figure}[!tbp]
\includegraphics[width=3.375in]{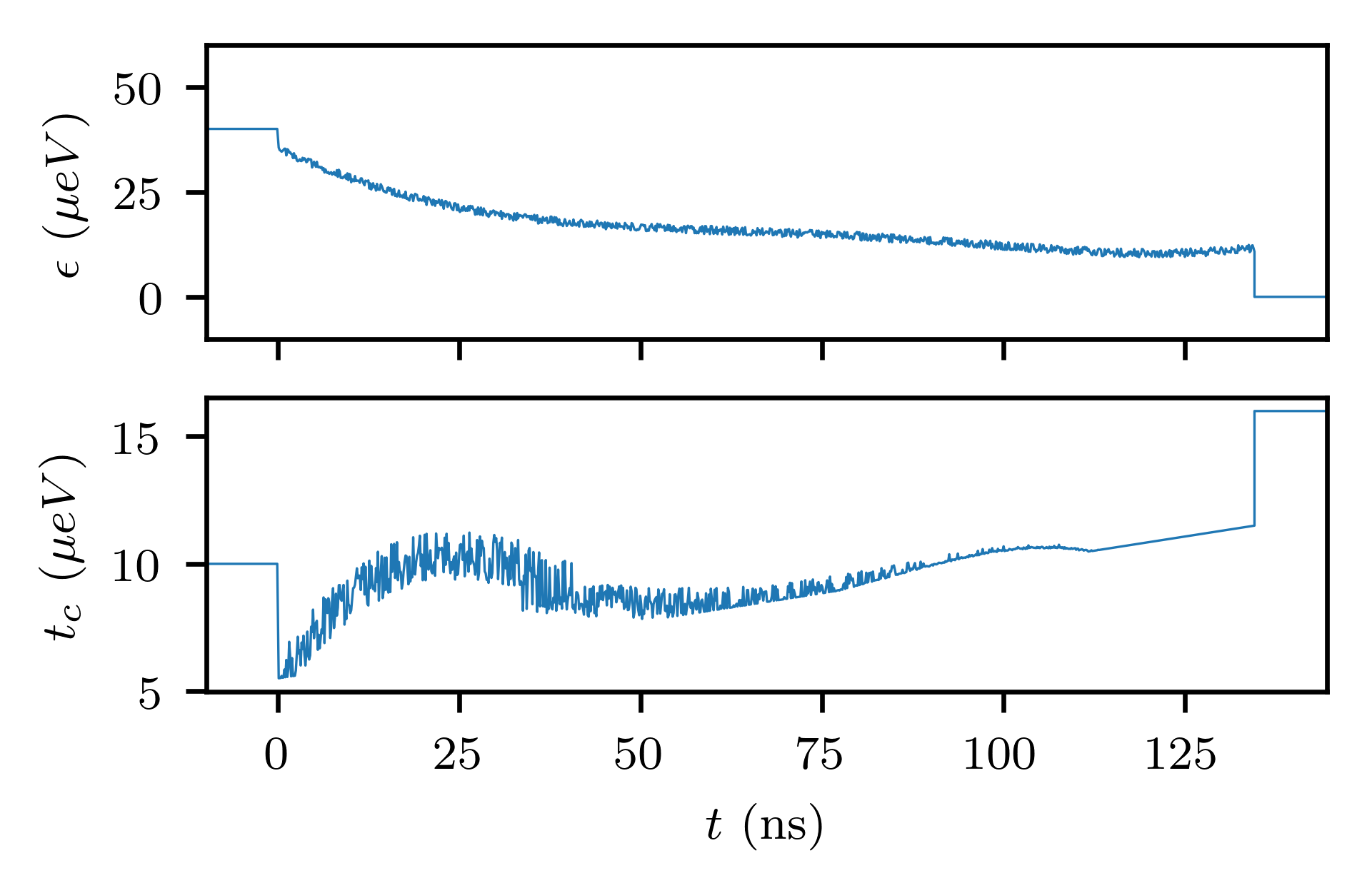}
\caption{Optimal $134.6$~ns control pulse generated by GRAPE for arbitrary state transfer. Electrostatic energies for $t<0$ and $t>T$ correspond to the initial memory-mode and final flopping-mode configurations, respectively.}
\label{fig:opt_ctrl_pulse}
\end{figure}
In the context of a general quantum information processing sequence, the qubit state cannot be assumed to be an eigenstate. In this case, optimal control pulses do not significantly accelerate high-fidelity operating regime transfers; however, maximal fidelity is increased (Fig. \ref{fig:arbitrary_tx_fid}).
In the case of completely unknown qubit states, simulated by applying the control pulses to logic states of the general form $\ket{\psi} = \alpha \ket{0} + \beta \ket{1}$ with $\alpha$, $\beta$ randomly generated complex numbers, it is found that peak fidelity is always reached with $134.6$ ns control pulses, showing that the operator approach to optimal pulse design yields state-agnostic controls (Fig. \ref{fig:random_tx_fid}), which is not the case in general for linearly ramped controls. 
When the optimal $134.6$ ns control pulse is applied to a large number of randomly generated logic states, it is found that this approach designs a state-robust pulse, with high fidelity, low variance operating regime transfers, which do not require a priori knowledge of the qubit's state (Fig. \ref{fig:random_tx_stats}). This is crucial for quantum computation applications, as the qubit's operating regime will need to be transferred several times during a computation, without having access to the qubit's state.

\section{Experimental Considerations}
\label{sec:experiment}
\begin{figure}[!tbp]
\includegraphics[width=3.375in]{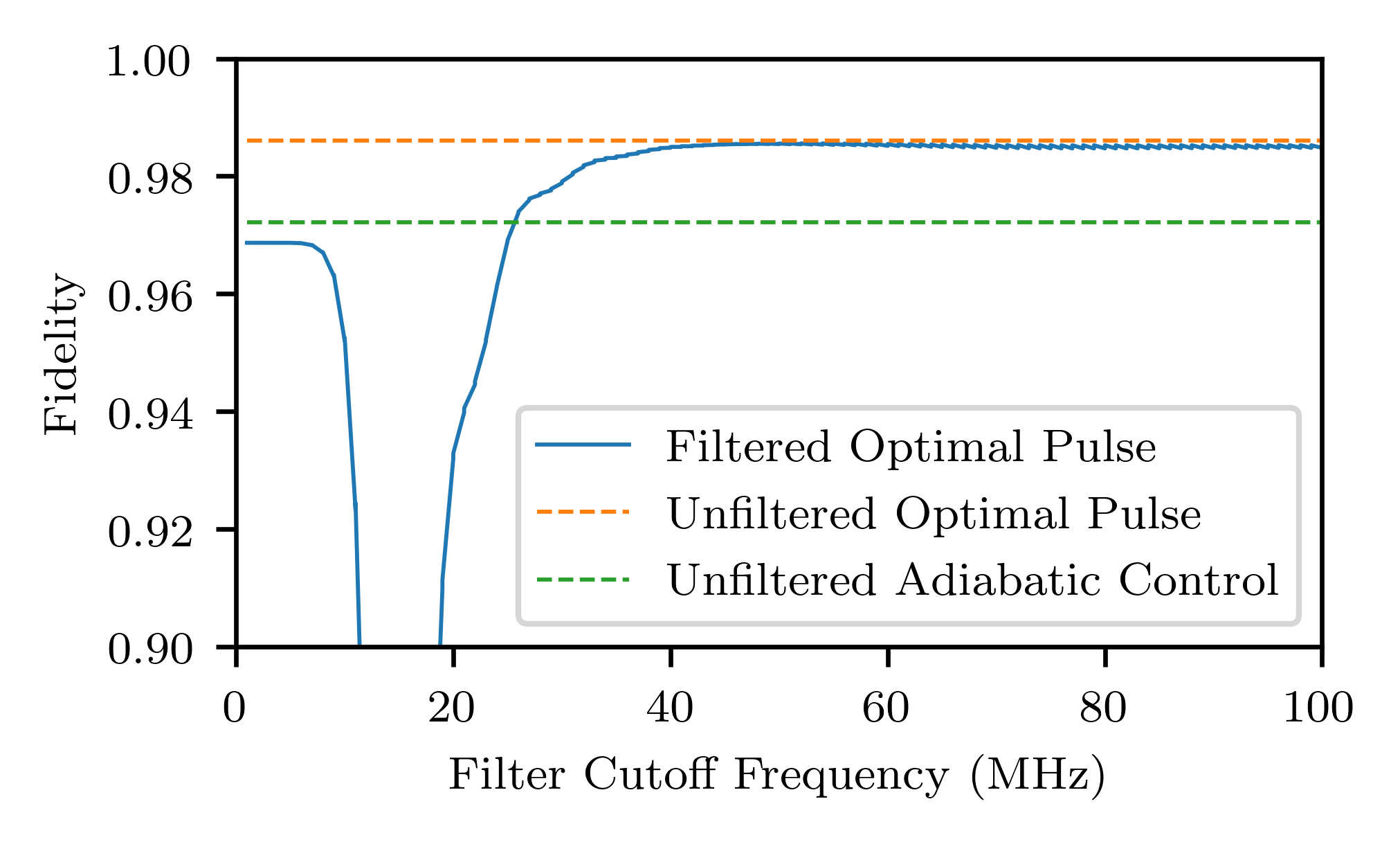}
\caption{Operating regime transfer fidelity from the initial memory-mode to the final flopping mode configuration for qubit state $\ket{0} + \ket{1}$, as a function of the low-pass filter cutoff frequency applied to the control signal. As expected, the transfer fidelity attained by the filtered control pulse (solid blue line) approaches that of adiabatic ramping (green dotted line) for very low cutoff frequencies, whereas higher cutoff frequencies approach unfiltered optimal control performance (orange dotted line).}
\label{fig:cutoff_vs_fidelity}
\end{figure}
\begin{figure}[!tbp]
\includegraphics[width=3.375in]{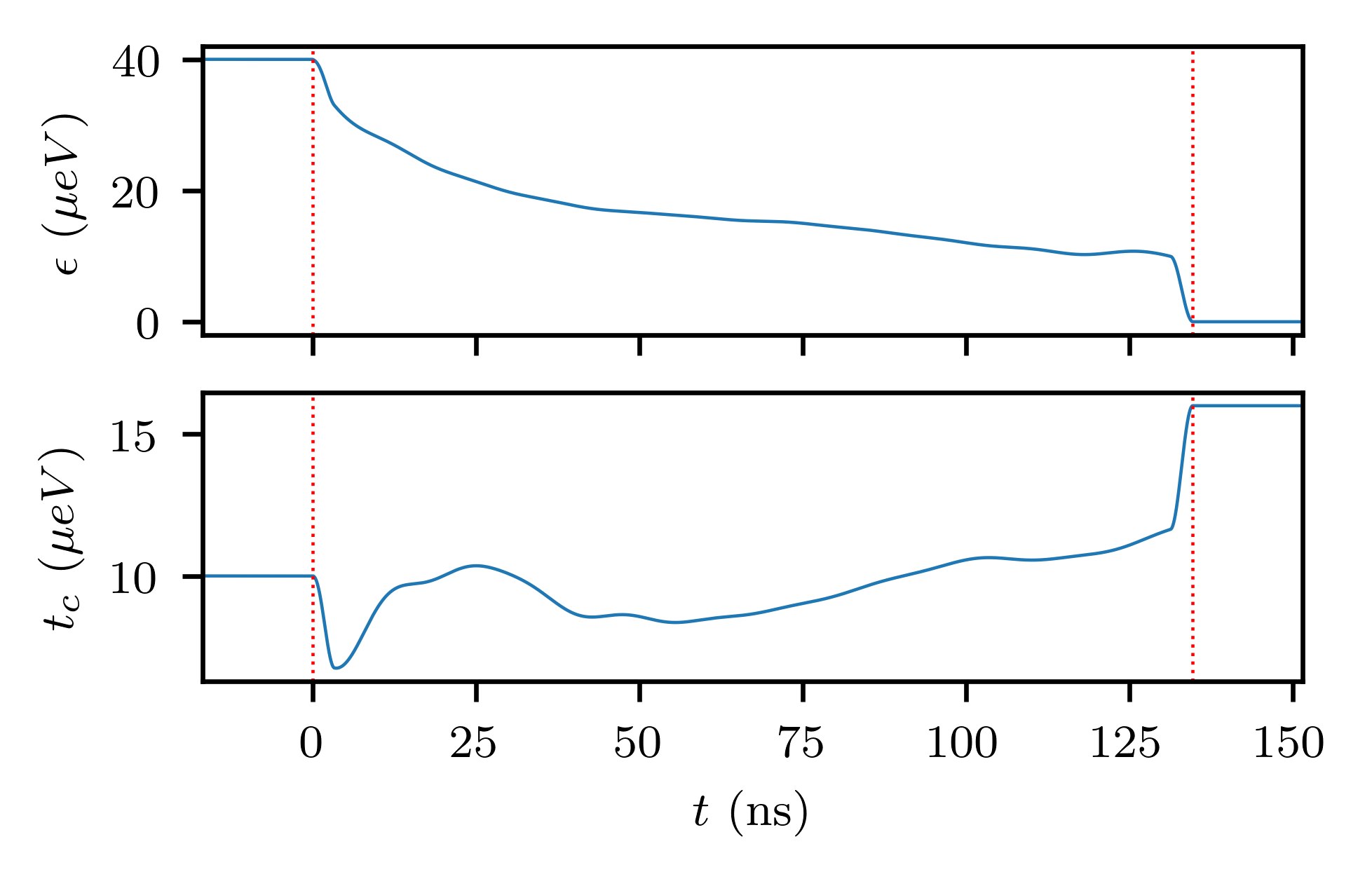}
\caption{Filtered and windowed $134.6$~ns control pulse. The optimal control pulse (Fig. \ref{fig:opt_ctrl_pulse}) was passed through a $80$~MHz lowpass filter and windowed to eliminate discontinuities at the boundaries. The vertical lines correspond to $t=0$ and $t=T$; electrostatic energies outside these bounds correspond to the initial memory-mode and final flopping mode configurations, respectively.}
\label{fig:filter_ctrl_pulse}
\end{figure}
\begin{figure}[!tbp]
\includegraphics[width=3.375in]{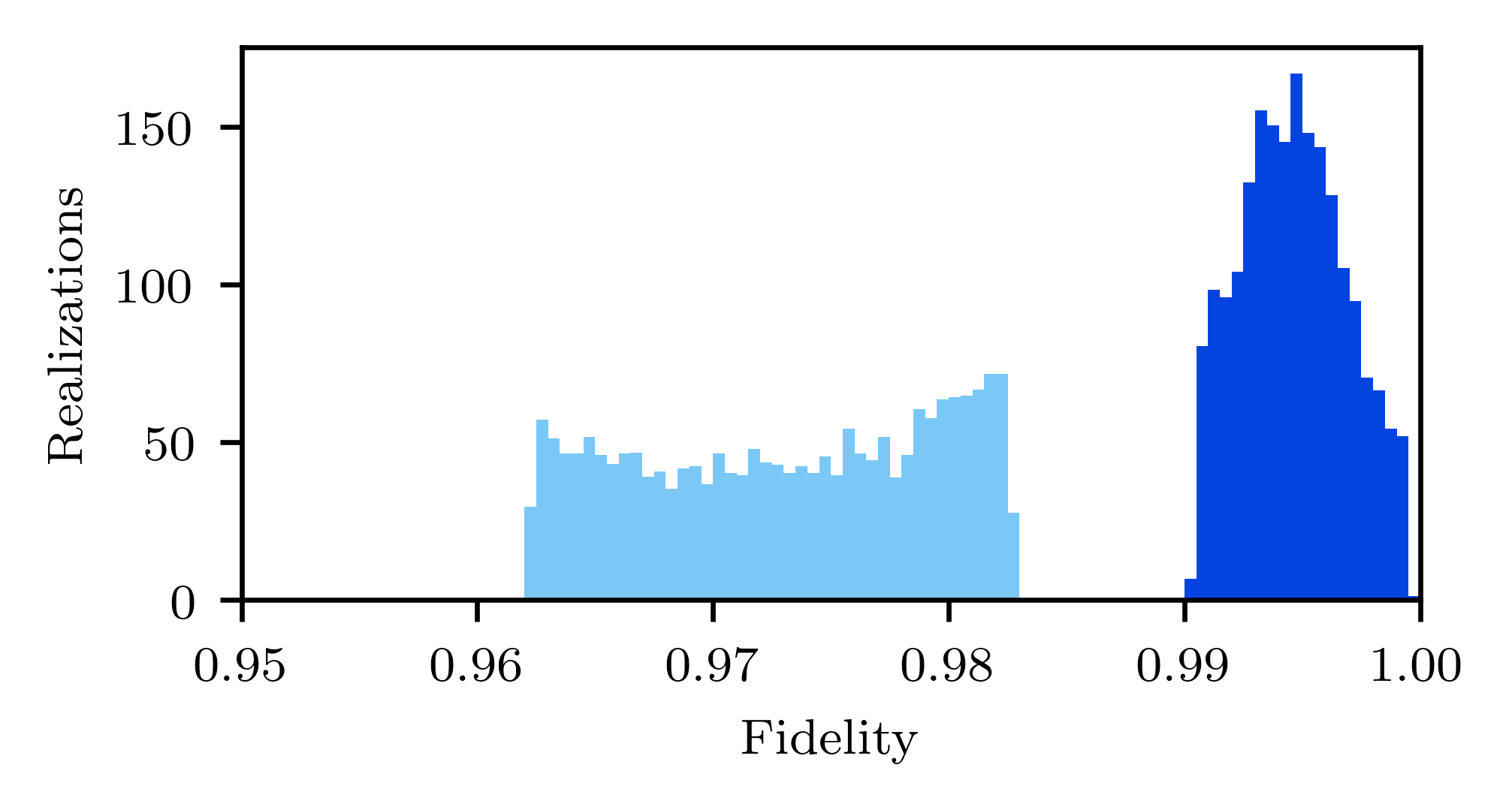}
\caption{Evolution statistics for 5000 randomly-generated logical states, using the filtered and windowed signal shown in Fig. \ref{fig:filter_ctrl_pulse}. The filtering and windowing operations lead to no significant decrease in control performance when compared to the results shown in Fig. \ref{fig:random_tx_stats}.}
\label{fig:filter_tx_stats}
\end{figure}
The resulting optimal pulse used to transfer arbitrary qubit states from the memory-mode to the flopping-mode operating regime is shown in Fig. \ref{fig:opt_ctrl_pulse}. As is common when using the GRAPE algorithm, the final pulse presents two undesirable features, namely high-frequency spectral content (up to tens of GHz) and sharp discontinuities at the beginning and at the end of the control. These features render experimental reproduction of these pulses very difficult even with modern arbitrary waveform generators, especially considering the low-pass filtering imposed by cryogenic control lines, despite previous successful implementations of GRAPE-designed control pulses in experiments~\citep{Nimbalkar2012, Groszkowski2011, Yang2020, Zong2021}.
The first of these undesirable features can be negated by using a simple low-pass filter on the optimal control pulse and observing the effects of a variable cutoff frequency on the resulting state fidelity, which are illustrated in Fig. \ref{fig:cutoff_vs_fidelity}. As shown, there is no appreciable decay in the quality of the state transfer even for cutoff frequencies as low as 100~MHz, much lower than the maximum bandwitdh on high-frequency control lines of cryogenic apparatus.
The second undesirable feature (sharp discontinuities) is removed by applying a Tukey window with a small $\alpha$-factor (here $\alpha = 0.05$), effectively forcing initial and final control pulse amplitudes to the steady-state values for $\epsilon$ and $t_c$ in the initial and final operating regime, respectively.
The resulting filtered and windowed control pulse is shown in Fig. \ref{fig:filter_ctrl_pulse}. Once again, this experimentally feasible version of the optimal control pulse has been applied in simulation to randomly-generated logical states of the general form $\ket{\psi} = \alpha \ket{0} + \beta \ket{1}$ with $\alpha$, $\beta$ randomly generated complex numbers, showing virtually no loss of fidelity from the low-pass filtering and windowing of the control signal (see Fig. \ref{fig:filter_tx_stats}).

\section{Conclusion}
\label{sec:conc}
It was shown that an algorithmic approach based on GRAPE to designing control pulses by iterating towards a target operator rather than a target state leads to state-robust control pulses for DQD operating regime transfers. When the qubit is prepared in a known Hamiltonian eigenstate, the operating regime transfer can be executed significantly faster than with a standard adiabatic pulse while reaching fidelities greater than $99\%$. When the qubit is prepared in any other arbitrary state, optimal control pulses which have been filtered and windowed for experimental feasibility reach fidelities greater than $99\%$ without a priori knowledge of the qubit state, which was not possible using linear ramping of the electrical parameters. 


\begin{acknowledgments}
VR acknowledges financial support via B1X~-~Bourses de maîtrise en recherche from the Fonds de recherche du Québec~-~Nature et technologies and Bourse VoiceAge pour l'excellence académique aux études supérieures from Université de Sherbrooke. YBL additionally acknowledges support from the Canada First Research Excellence Fund and Institut quantique at Université de Sherbrooke.
\end{acknowledgments}

\appendix*

\section{Charge Relaxation Rate}
\label{app:charge_relax}

The rate of charge relaxation via emission of longitudinal acoustic (LA) phonons in a steady-state regime can be obtained by writing the charge-phonon interaction Hamiltonian and using Fermi's golden rule at 0~K, yielding~\cite{Benito2017}
\begin{equation}
\gamma_{1,c} = \frac{2\pi}{\hbar} \sum_{\vec{k}} \lvert c_{\vec{k}} \rvert^2 \delta(\Omega-E_f) = \frac{2\pi}{\hbar} \lvert c_k \rvert^2 D(\Omega),
\end{equation}
with phonon energy $E_f$. This equation shows an implicit dependence of the charge relaxation rate on the orbital energy $\Omega$, which we now make explicit by using the Debye model for the phonon density of states~\cite{Kittel2004}
\begin{equation}
\label{eq:debye-model}
D(\Omega) = \frac{V\Omega^2}{2\pi^2\hbar^2\nu_s^3},
\end{equation}
and the expression for the coefficient $c_k$~\cite{Zhao2008}
\begin{equation}
\label{eq:phonon-coefficient}
\lvert c_k \rvert^2 = \frac{d^2}{2\rho\nu_s^2V} \hbar \nu_s k.
\end{equation}
In Eqs. (\ref{eq:debye-model}) and (\ref{eq:phonon-coefficient}), $d$ is the deformation potential, $V$ is the device volume, $\Omega$ is the orbital energy, $\nu_s$ is the speed of sound in the crystal, $\rho$ is the crystal density and $k$ is the wave vector modulus. Using the dispersion relation
\begin{equation}
\label{eq:dispersion-relation}
k = \frac{2\pi}{\lambda} = \frac{\Omega}{\hbar\nu_s},
\end{equation}
the charge relaxation rate can be written in energy units as
\begin{equation}
\label{eq:charge-relaxation-energy}
\gamma_{1,c} = \frac{1}{2\pi\hbar^3}\frac{\Omega^3 d^2}{\rho\nu_s^5}.
\end{equation}
It is thus seen that the charge relaxation rate via emission of LA phonons has cubic dependence on the orbital energy of the qubit, and therefore depends on the amplitude of the control voltages applied to the gates. Using $d = 8.89$~eV, $\rho = 2330$~kg/m\textsuperscript{3} and $\nu_s = 8433$~m/s\textsuperscript{2} for silicon~\citep{Witzens2014, Sham1963, Hopcroft2010}, Eq. (\ref{eq:charge-relaxation-energy}) yields charge relaxation rates which are consistent with experimental measurements in literature~\citep{Mi2018}.

However, in the context of controlling the operating regime of the DQD, charge relaxation via emission of LA phonons in the silicon lattice is dependent on the control amplitudes $\epsilon$ and $t_c$ via the electron orbital energy $\Omega \equiv \Omega(\epsilon, t_c)$. Attempting to optimize the control signals on $\epsilon$ and $t_c$ while taking the exact dependence of this relaxation process on $\Omega(\epsilon, t_c)$ leads to transcendental equations which are extremely computationally expensive to solve, for a very small projected gain in precision given the short duration of the control sequence relative to the decoherence rates involved. Instead, constant rates chosen to represent the worst-case given the dynamic range for parameters $\epsilon$ and $t_c$ are used in simulation.

\bibliography{references}

\end{document}